\documentclass[superscriptaddress,showpacs,preprintnumbers,amsmath,amssymb,
nofootinbib,twocolumn,aps,prd,10pt]{revtex4-1}
\usepackage{amssymb,amsmath}
\usepackage{epsfig}
\usepackage{xcolor}
\usepackage[utf8]{inputenc}
\usepackage{stmaryrd}
\usepackage{mathrsfs}
\usepackage{mathalfa}
\usepackage[normalem]{ulem}

\newcommand{\be}{\begin{equation}}
\newcommand{\ee}{\end{equation}}
\newcommand{\ba}{\begin{eqnarray}}
\newcommand{\ea}{\end{eqnarray}}
\newcommand{\beg}{\begin{gather*}}
\newcommand{\eng}{\end{gather*}}
\newcommand{\hh}{,\hspace{0.5cm}}
\newcommand{\hhh}{,\hspace{0.2cm}}

\newcommand{\n}[1]{\label{#1}}

\newcommand{\ts}[1]{{\boldsymbol{#1}}}

\newcommand{\p}[1]{\ts{\pi}^{(#1)}}
\newcommand{\pb}[1]{\ts{\bar{\pi}}^{(#1)}}
\newcommand{\ps}[1]{\ts{\pi}^{s (#1)}}

\newcommand{\dl}{\nabla_{\ts{l}}}

\newcommand{\as}{\ts{\cal A}^{s}}
\newcommand{\fs}{\ts{\cal F}^{s}}

\newcommand{\rav}[1]{\stackrel{#1}{=}}

\def\XXint#1#2#3{{\setbox0=\hbox{$#1{#2#3}{\int}$ }
\vcenter{\hbox{$#2#3$ }}\kern-.6\wd0}}

\usepackage[makeroom]{cancel}
\usepackage[caption=false]{subfig}
\usepackage[colorlinks=true,
            citecolor=green,
            linkcolor=red,
            filecolor=cyan,
            urlcolor=magenta,
            backref=false]{hyperref}

\begin{document}

\title{Maxwell equations in a curved spacetime: Spin optics approximation}

\author{Valeri P. Frolov}
\email{vfrolov@ualberta.ca}
\affiliation{Theoretical Physics Institute, University of Alberta, Edmonton, Alberta, Canada T6G 2E1}


\begin{abstract}
We study propagation of high-frequency electromagnetic waves in a curved spacetime. We demonstrate how a modification of the standard geometric optics allows one to include the helicity dependent corrections into the equations of motion of circularly polarized beams of radiation. As a result, polarized light rays are still null but not geodesic curves. To achieve these results we  construct null frames associated with a set of (non-geodesic) null rays and use these frames for description of the high-frequency wave propagation. We call this approach spin optics approximation. It is completely covariant and it can be used in an arbitrary time-dependent gravitational field.
\end{abstract}


\maketitle

\section{Introduction}

Short wave (or high-frequency) approach is a powerful method of construction of approximate solutions of linear differential equations with spatially varying coefficients. In application to the linear partial differential equations it allows one to find their asymptotic solutions by reducing this problem to studying of hamiltonian dynamical systems. This method is widely used in different areas of physics and has different (historically motivated) names. In quantum mechanics this method is known as a quasiclassical or WKB approximation, by initials for Wentzel-Kramers-Brillouin. In wave optics this method is known as a geometric optics approximation. It takes its origin with the paper by Debay in 1911 \cite{Debay:1911} and was developed in many subsequent publications (see e.g. \cite{born:1999} and references therein).

The main idea of this approach is to search for a solution $\Phi$ of the wave equation in the form $\Phi\approx f \exp{iS}$, where the phase function $S$ rapidly changes, while the field amplitude $f$ is a slowly varying function. This means that $\nabla S\sim \omega$, where $\omega$ is a characteristic large frequency. Substituting this ansatz into the field equations and keeping the terms of the lowest order in $1/\omega$   one obtains a first order partial differential equation of the form $H(\nabla S,x)=0$ known as an eikonal equation. This equation can be identified with the Hamilton-Jacobi equation. Putting $P=\nabla S$ one reduces the problem to study of the corresponding  dynamical system with Hamiltonian $H(P,x)$. In the application to the standard optics, the equation $S=$const determines a wavefront and the vectors $\ts{P}$ orthogonal to the wave front are tangent to the light rays. In the phase space a beam of such null rays forms a Lagrangian submanifold (for details see e.g. a remarkable book \cite{Arnold:1989}).

Geometric optics approximation for obtaining high-frequency asymptotic solutions of Maxwell equations in a curved spacetime has been discussed in a number of publications starting with the papers \cite{Sachs:1961zz,Robinson:1961,Kristian:1965sz,Ehlers:1967}.  A remarkable summary of the geometric optics in a curved spacetime can be found in the book \cite{Misner:1974qy} (see also more recent papers  \cite{Seitz:1994xf,Dolan:2018nzc,Dolan:2018ydp} and references therein).
In a curved spacetime in the leading order of the geometric optics approximation,  light rays are null geodesics and the wave polarization vector is parallel-propagated along the rays. For a beam of light the square of its amplitude is inverse proportional to the area of the cross-section of the beam. Parallel transport of the polarization vector of the electromagnetic wave results in the gravitational Faraday rotation effect \cite{Plebanski:1960,Piran:1985,Ishihara:1987dv,Nouri_Zonoz:1999,Sereno:2004jx}. Similar Faraday rotation effect exists also for polarized high-frequency gravitational waves propagating  in a curved spacetime \cite{Hou:2019wdg}. This effect is an analogue of the well-known electromagnetic Faraday effect for light propagating in magneto-active media \cite{Rytov:1938,Pershan:1967,Kravtsov:1990}.

Gravitational Faraday effect is one of manifestations of action of gravimagnetic forces on a particle with spin or spin-curvature interaction. Its dual is an effect of the helicity dependence of the motion of circularly polarized light beams in the gravitational field of a spinning objects \cite{Mashhoon:1974,Mashhoon:1975ki,Mashhoon:2008si}.
This is an analogue of a so called optical Magnus effect \cite{Magnus:1992,Bliokh:2004}, that is polarization-dependence of light propagation in an inhomogeneous optical media. This effect sometime is also called {\em Hall effect of light}
\cite{Onoda:2004}. Comprehensive discussion of this subject and numerous references can be found in recent reviews \cite{SPINHALL:2015,SPINHALL:2017}.

There are many publications in which different approaches to the gravitational spin Hall effect were proposed. They can be divided into three main groups. (i) Adaptation of the Mathisson-Papapertou-Dixon equations for the motion of massive particles with spin to the massless case (see e.g. \cite{Souriau:1974,Saturnini:1976,Duval:2008,Duval:2019}); (ii) Using the methods of quantum mechanics, such as Foldy-Wouthuysen transformation and Berry phases \cite{Gosselin:2007};
(iii) Spin optics or modified geometric optics \cite{Frolov:2011mh,Frolov:2012zn,Yoo:2012vv,Shoom:2020zhr}.
In the latter approach the standard geometric optics eikonal function is modified by including into it a specially chosen helicity dependent correction. This correction contains extra factor $\omega^{-1}$ and hence
is suppressed at very high-frequencies. However, at finite frequency and for travel of polarized rays at large distances they can modify the ray propagation and become important. Comprehensive discussion and comparison of different approaches to the gravitational spin Hall effect can be found in a recent review \cite{Oancea:2019pgm}.

Let us emphasize that development of the high-frequency approximation for propagation of the field with spin in a curved background has a well known problem.  For a many component field there is an ambiguity: one can make a phase rotation of the multi-component prefactor amplitude and compensate this by the change of the phase function $S$ \cite{Weinberg:1962,Bjorken:1981,FUK:1996}. A similar problem arises in the application of the WKB approximation to an equation for a particle with spin moving in an inhomogeneous magnetic field. Since the spin is proportional to $\hbar$, in the WKB limit $\hbar\to 0$ the spin contribution to the  motion formally vanishes \cite{PAULI:1932}. However, in such a version of the Stern-Gerlach experiment an observer registers  a  deflection of particles with different orientation of the spin by fixing points where the particles struck a detector screen. A key role in the adaptation of the WKB approximation to such a case is understanding that besides the length scale $\ell$ connected with the inhomogeneity of the magnetic field  there is another length parameter, a distance to the screen $L$, which can be much larger than $\ell$. To catch hold of this effect one should first diagonalize the corresponding Pauli equation, and then "enhance" the spin dependent terms by including them into the eikonal \cite{HSU:2011}.

A similar basic idea was used in the spin optics approach \cite{Frolov:2011mh,Frolov:2012zn}. For Maxwell theory it is possible to split the wave solutions into right and left-handed circularly polarized modes. The electromagnetic field with right and left hand polarization can be identified with self and anti-self dual solutions of the Maxwell equations, respectively. In spin optics one uses the WKB ansatz for each of these  solutions and does not require that the eikonal functions are the same for both types of waves. Instead of this, one includes into the eikonal the first order  high-frequency correction which is sensitive to the helicity. This method, called spin optics approximation, was developed in \cite{Frolov:2011mh,Frolov:2012zn} for polarized light propagation in a stationary spacetime. During some period of time it remained unclear how to spread  this approach to a  case when the gravitational field is not stationary. Recently this problem was solved. This remarkable breakthrough was achieved in paper \cite{Oancea:2020khc}. In the present paper we present a slightly different approach to this problem.

We assume that a high-frequency self or anti-self dual solution of the Maxwell equations is associated with the congruence of null rays which, in a general case, is not geodesic. In section~II we introduce the notion of $F$-transport which allows one to determine a null frame associated with a given set of null rays. In section~III we write Maxwell equations by using their high-frequency decomposition and obtain a set of truncated equations by keeping zero and first order terms in this system. We also use the null frames constructed in section~II to derive constraints on the field that follow from the property of its  self or anti-self-duality. In section IV we reproduce the results of the standard geometric optics approximation for the obtained truncated system of equations. The spin optics approach to the problem of the high-frequency polarized light propagation in a curved spacetime is developed in section~V. In this section we  derive the corresponding equations of motion for polarized light rays. These equations determine the acceleration of the worldlines of null rays and hence specify the choice of the null ray congruence. The ray equations together with equations for null frame propagation along null rays give a complete self-consistent set of equations. In section~VI we discuss an effective action for polarized light rays and compare Lagrangian and Hamiltonian formulations for this problem. Final section~VI contains brief summary of the obtained results and their discussion. In this paper we use the signs convention adopted in the book \cite{Misner:1974qy}. In particular, the metric has the signature $(-,+,+,+)$.
In description of vectors and tensors we use both their coordinate and coordinate-free forms. In the latter case we denote these objects by  boldface letters. For example, a scalar product of two vectors $\ts{a}$ and $\ts{b}$ is
$(\ts{a},\ts{b})=g_{\mu\nu}a^{\mu} b^{\nu}$. We also denote $\ts{a}^2=(\ts{a},\ts{a})$.

\section{Null tetrads and polarization tensors}
\n{NFR}
\subsection{Null tetrads}

\n{NF}

Let us consider a congruence of null curves which later will be identified with trajectories of  a massless particle with the spin. Let $\ts{l}$ be tangent vectors to these null curves, and we denote
\be
w^{\mu}=l^{\nu}l^{\mu}_{\ ;\nu}\, .
\ee
Since $\ts{l}^2=0$ one has
\be
(\ts{l},\ts{w})=l_{\mu}w^{\mu}=0\, .
\ee
Let us consider integral lines $x^{\mu}(\lambda)$ of $\ts{l}$
\be
{dx^{\mu}(\lambda)\over d\lambda}=l^{\mu}\, .
\ee

We complement the vector field $\ts{l}$ by three other null fields $\ts{n}$, $\ts{m}$ and $\ts{\bar{m}}$ and require that the following  normalization conditions are satisfied
\be\n{norm}
(\ts{l},\ts{n})=-1, \quad (\ts{m},\ts{\bar{m}})=1\, ,
\ee
while all other scalar products of these vectors vanish. We also assume that this frame has a right-handed orientation.
In what follows we shall widely use this (complex) null tetrad $\{\ts{l},\ts{m},\ts{\bar{m}},\ts{n}\}$.

For a fixed congruence of null rays there exist a following freedom in the choice of the vectors of the null tetrad:
\begin{enumerate}
\item $\ts{l} \to A \ts{l}$, \hh $\ts{n} \to A^{-1} \ts{n}$ ;
\item $\ts{l} \to \ts{l}$, \hhh $\ts{m} \to \ts{m}+a\ts{l}$, \hhh
$\ts{\bar{m}} \to \ts{\bar{m}}+\bar{a}\ts{l}$, \newline
$\ts{n}\to \ts{n}+\bar{a}\ts{m}+a \ts{\bar{m}}+a \bar{a}\ts{l}$
;
\item  $\ts{m} \to e^{i\varphi}\ts{m}$\hh $\ts{\bar{m}} \to e^{-i\varphi}\ts{\bar{m}}$.
\end{enumerate}
Here $a$ is a complex function, and $A$ and $\varphi$ are two real functions. The first of these transformations reflects a freedom of the choice of the parametrization along null rays. Under this transformation, the vector $\ts{w}$ changes as follows
\be\n{AA}
\ts{w}\to A^2\ts{w}+A\nabla_l A \ \ts{l}\, .
\ee
The vector $\ts{w}$ is invariant under the transformations 2 and 3.

We demonstrate now how one can reduce the freedom of the transformations 1--3 and construct a special null frame associated with a given congruence of null rays. Consider a null ray $x^{\mu}(\lambda)$ and choose a vector $\ts{n}$ on this curve such that at $\lambda=\lambda_0$ it obeys the relations
\be \n{cond}
\ts{n}^2|_{\lambda_0}=0\hh (\ts{n},\ts{l})|_{\lambda_0}=-1\, .
\ee
We define its transport along the ray by the following equation
\be
\nabla_{\ts{l}}\ts{n}=(\ts{w},\ts{n}) \ts{n}\, .
\ee
Then one has
\ba
&&{d \over d\lambda}\ts{n}^2=2(\ts{w},\ts{n}) \ts{n}^2\, ,\\
&&{d \over d\lambda}[(\ts{n},\ts{l})+1]=(\ts{w},\ts{n})[(\ts{n},\ts{l})+1]\, .
\ea
Both of these equations are of the form ${dz/ d\lambda}=(\ts{w},\ts{n})z$. A solution of such an equation which vanishes at $\lambda_0$ is identically zero along the ray. This means that if one imposes  conditions (\ref{cond}) at the initial value $\lambda=\lambda_0$, then this vector $\ts{n}$ has the property $\ts{n}^2=0$ and $(\ts{n},\ts{l})=-1$ valid everywhere on the ray.

We use this vector to define a following tensor
\be
{\cal V}^{\mu}_{\ \nu}=w^{\mu}n_{\nu}-n^{\mu}w_{\nu}\, .
\ee
We introduce  operator ${\cal D}$ acting on a tensor field $A^{\mu\ldots}_{\ \ \nu\ldots}$ along the ray by the following relation
\be
{\cal D}A^{\mu\ldots}_{\ \ \nu\ldots}=\nabla_{\ts{l}}A^{\mu\ldots}_{\ \ \nu\ldots}+{\cal V}^{\mu}_{\ \lambda} A^{\lambda\ldots}_{\ \ \nu\ldots}+\ldots -{\cal V}^{\lambda}_{\ \nu}A^{\mu\ldots}_{\ \ \lambda \ldots}+\ldots\, .
\ee
Here $\nabla_{\ts{l}}$ is a covariant derivative along the vector field $\ts{l}$.
The operator  ${\cal D}$ when applied to a product of two tensors satisfies the Leibniz rule.
It is easy to check that ${\cal D} \ts{g}=0$ and the scalar product of any two vectors $\ts{a}$ and $\ts{b}$ is constant along the ray provided
${\cal D}\ts{a}={\cal D}\ts{b}=0$. The operator ${\cal D}$ may be considered as a modification of the Fermi derivative adapted to the case of null curves. We say that a tensor is $F$-propagated along ray if its ${\cal D}$-derivative  vanishes. In particular, for a $F$-propagated vector $\ts{z}$ one has
\be
\dl\ts{z}=(\ts{w},\ts{z}) \ts{n}-(\ts{n},\ts{z}) \ts{w}\, .
\ee
By construction the vector $\ts{l}$ is $F$-propagated along the ray.

We choose complex null vectors $\ts{m}$ and $\bar{\ts{m}}$ to be $F$-transported along the ray and such that at the initial point of the ray $\lambda_0$ they obey the relations
\ba
&&(\ts{m},\ts{l})|_{\lambda_0}=(\bar{\ts{m}},\ts{l})|_{\lambda_0}=
(\ts{m},\ts{n})|_{\lambda_0}=(\bar{\ts{m}},\ts{n})|_{\lambda_0}=0\, ,\nonumber\\
&&(\ts{m},\ts{m})|_{\lambda_0}=
(\bar{\ts{m}},\bar{\ts{m}})|_{\lambda_0}=0\hh
(\ts{m},\bar{\ts{m}})|_{\lambda_0}=1\, .
\ea
Since $F$-transport preserves the scalar product a so defined null frame $(\ts{l},\ts{m},\bar{\ts{m}},\ts{n})$ obeys the normalization conditions (\ref{norm}) along the ray and satisfies the equations
\ba
&& \dl\ts{n}=(\ts{w},\ts{n}) \ts{n}\, ,\nonumber\\
&& \dl\ts{m}=(\ts{w},\ts{m}) \ts{n}\, ,\n{nmm}\\
&& \dl\ts{\bar{m}}=(\ts{w},\ts{\bar{m}}) \ts{n}\, .\nonumber
\ea
Let us emphasize that after imposing $F$-transport requirement on the null frame, the freedom 2 and 3 in its choice  is reduced by the conditions $\dl a=\dl\varphi=0$.

Equations (\ref{nmm}) can be further simplified. Namely, one can always choose the function $A$ in (\ref{AA}) so that $(\ts{w},\ts{n})=0$. The only ambiguity which is left in the transformation 1 is $A=$const along the rays. This condition fixes the choice of the parameter $\lambda$ along the ray up to its possible  rescaling $\lambda\to A^{-1}\lambda$, where $A$ is constant.
In what follows we always use this parametrization and call it {\em{canonical}}. In the canonical parametrization the vector $\ts{n}$ is parallel transported along the null rays, $\dl\ts{n}=0$, and the vector $\ts{w}$ is of the form
\be\n{www}
\ts{w}=-\bar{\kappa} \ts{m}-\kappa \ts{\bar{m}}\hh\kappa=-m^{\mu} l^{\nu}l_{\mu ;\nu}\, .
\ee

Let us summarize: In the canonical parametrization the $F$-transported null tetrad vectors obey the equations
\be\n{NNN}
  \dl \ts{n}=0\hhh \dl \ts{m}=-\kappa \ts{n}\hhh \dl \ts{\bar{m}}=-\bar{\kappa} \ts{n}\, ,
\ee
and the remaining freedom of the tetrad transformation  1-3  is reduced to such transformations performed on the null tetrad vectors at some initial moment of time.
We call a so defined $F$-propagated tetrad a {\em null frame associated with the congruence of null rays}.

\subsection{Polarization tensors}

We denote by $\ts{e}$ a unit volume 4-form
\be
\ts{e}=i\ \ts{l}\wedge \ts{m}\wedge\ts{\bar{m}}\wedge\ts{n}\, .
\ee
Its coordinate form is $e_{\mu\nu\lambda\rho}=i 4! \ l_{[\mu}m_{\nu}\bar{m}_{\lambda}n_{\rho]}$ .
If $\ts{\omega}$ is a rank $p$-form then using this tensor one can define a Hodge dual $(4-p)$-form $\star\ts{\omega}$. In particular,  if a two-form $\ts{\omega}$ has coordinates $\omega_{\mu\nu}$, then $(\star\omega)_{\mu\nu}={1\over 2} e_{\mu\nu\lambda\rho}\omega^{\lambda\rho}$.

We introduce the following three two-forms $\p{a}$, $a=0,1,2$ by relations
\be\n{polt}
\p{0}=\ts{\bar{m}}\wedge\ts{n}\, ,\
\p{1}=-(\ts{l}\wedge\ts{n}-\ts{m}\wedge\ts{\bar{m}})\, ,\
\p{2}=\ts{l}\wedge\ts{m}\, .
\ee
In the coordinate form the components of these two-forms are
\ba
&&\pi^{(0)}_{\ \mu\nu}=2\bar{m}_{[\mu} n_{\nu]}\hhh
\pi^{(2)}_{\ \mu\nu}=2l_{[\mu} m_{\nu]}\, ,\nonumber\\
&&\pi^{(1)}_{\ \mu\nu}=2\left(-{l}_{[\mu} n_{\nu]}+{m}_{[\mu}\bar{m}_{\nu]}\right)\, .
\ea

We denote by $\pb{a}$  two-forms obtained from $\p{a}$ by their complex conjugation.
These forms $\p{a}$ and $\pb{a}$ obey the property
\be
\star\p{a}=i\p{a}\hh \star\pb{a}=-i\pb{a}\, .
\ee
In other words, the forms $\p{a}$ are self-dual, while  $\pb{a}$ are anti self-dual.
We call these objects polarization tensors.

Let $\ts{z}$ be a vector, then we denote by $\p{a}\cdot \ts{z}$ a vector with components
$\pi^{(a)\mu}_{\ \ \ \ \ \nu}z^{\nu}$. Then the action of the polarization tensors on the null-tetrad vectors can be written in the form
\ba
&&\p{0}\cdot \left(\begin{array}{c}
                   \ts{l} \\
                   \ts{m} \\
                   \ts{\bar{m}} \\
                  \ts{n}
                 \end{array}
                 \right)=\left(\begin{array}{c}
                  -\ts{\bar{m}} \\
                  -\ts{n} \\
                   0 \\
                  0
                 \end{array}
                 \right)\,  ,\nonumber\\
&&\p{1}\cdot \left(\begin{array}{c}
                   \ts{l} \\
                   \ts{m} \\
                   \ts{\bar{m}} \\
                  \ts{n}
                 \end{array}
                 \right)= \left(\begin{array}{c}
                   \ts{l} \\
                   \ts{m} \\
                   -\ts{\bar{m}} \\
                  -\ts{n}
                 \end{array}
                 \right)\, ,\\
&&\p{2}\cdot \left(\begin{array}{c}
                 \ts{l} \\
                   \ts{m} \\
                   \ts{\bar{m}} \\
                  \ts{n}
                 \end{array}
                 \right)= \left(\begin{array}{c}
                   0 \\
                   0 \\
                   \ts{l} \\
                  \ts{m}
                 \end{array}
                 \right)\,  .\nonumber
\ea

We define a contraction $\ts{a}\circ \ts{b}$ of 2 two-forms $\ts{a}$ and $\ts{b}$ as follows
\be
\ts{a}\circ \ts{b}=\ts{b}\circ \ts{a}= {1\over 2}a_{\mu\nu} b^{\mu\nu}\, .
\ee
It is easy to check that
\ba
&&\pb{a}\circ \p{b}=0\, ,\n{ppp}\\
&&\p{a}\circ \p{b}=\pb{a}\circ \pb{b}=
\delta_0^a\delta_2^b+\delta_2^a\delta_0^b-2\delta_1^a\delta_1^b\, .\n{ddd}
\ea
It is convenient to combine $\p{a}$ and $\pb{a}$ into a unique set of two-forms $\ts{\pi}^{s (a)}$ by specifying its components as follows
\be \n{ps}
\ts{\pi}^{+1 (a)}=\p{a}\hh \ts{\pi}^{-1 (a)}=\pb{a}\, .
\ee
We call $s=\pm 1$ a helicity parameter.
The forms $\ts{\pi}^{s (a)}$ form a basis in a six-dimensional linear space of two-forms.

\section{Self-dual and anti-self-dual solutions of Maxwell equations}

\subsection{Field equations}

In the absence of currents the Maxwell equations for the electromagnetic field $\ts{F}$ have the following standard form
\be
d\ts{F}=\delta\ts{F}=0\, ,
\ee
where a co-derivative $\delta$ is defined as $\delta=\star d \star$. We denote
\be
\ts{{\cal F}}^{s}={1\over 2} [\ts{F}-is (\star\ts{F})]\, ,
\ee
where $s=\pm 1$. In the four dimensional  spacetime with the Lorentz signature the Hodge duality operator has the property
$\star\star \ts{F}=-\ts{F}$, so that $\star \ts{\cal F}^{s}=is \ts{{\cal F}}^{s}$ .
Hence, the field $\ts{\cal F}^{+1}$ is self-dual, while $\ts{\cal F}^{-1}$ is anti-self-dual.

We consider $\ts{\cal F}^{s}$ as two independent complex fields and identify the parameter $s$ with the helicity of the field. These fields obey the equations
\be
d\ts{\cal F}=\delta\ts{\cal F}=0\,  .
\ee

Using   relations (\ref{ppp})  one can show that  the field $\fs$ obeys the relations
\be\n{pol}
\fs \circ \ts{\pi}^{-s (a)}=0\,  ,
\ee
and it can be presented in the form
\be
\fs=\sum_{a=1}^{3}\Phi^{s}_{a}\ps{a}\, .
\ee
Using relations (\ref{ddd}) one gets
\be
\ps{b}\circ \fs=\Phi^{s}_{0}\delta_2^b+\Phi^{s}_{2}\delta_0^b-2\Phi^{s}_{1}\delta_1^b\, .
\ee
Hence
\ba
&&\Phi^{s}_{0}=\ps{2}\circ \fs\hh \Phi^{s}_{2}=\ps{0}\circ \fs\, ,\\
&&\Phi^{s}_{1}=-{1\over 2}\ps{1}\circ \fs\, .
 \ea
If $\ts{\cal F}$ is a self-dual field then its complex components
\ba
&&\Phi_0={\cal F}_{\mu\nu}l^{\mu}m^{\nu}\hh \Phi_2={\cal F}_{\mu\nu}\bar{m}^{\mu}n^{\nu}\,  ,\nonumber\\
&&\Phi_1={1\over 2}{\cal F}_{\mu\nu}(l^{\mu}n^{\nu}+\bar{m}^{\mu}m^{\nu})\, .
\ea
coincide with the  standard  complex tetrad components of the electromagnetic field introduced by Teukolsky in his paper \cite{Teukolsky:1973}.

We denote by $\ts{\cal A}^{s}$ a complex vector potential such that
\be
\fs=d\as \, ,
\ee
and use the gauge freedom to impose the Lorentz condition
\be
\delta \as=0\, .
\ee

\subsection{High-frequency expansion}

\n{hfe}

We write a complex potential $\ts{\cal A}$ in the form
\be\n{anz}
\ts{\cal A}=\ts{a}\exp(iS)\, ,
\ee
To simplify the expressions we skip the helicity index $s$ both in the amplitude $\ts{a}$ and in the phase function $S$. We restore this parameter in the final results.
We assume that real function $S$ is a "fast changing" phase and write its gradient as follows
\be
S_{;\mu}=\omega p_{\mu}\, .
\ee
Since our goal is to construct asymptotic solutions of the Maxwell equations in the high-frequency approximation we assume that the frequency $\omega$ is large. In what follows we shall use $1/\omega$ expansion. In fact, if $l$ is a characteristic scale involved in the problem (such as the curvature of the wave front, the size and duration of the radiation beam and the radius of the spacetime curvature) then the small dimensionless parameter of expansion is $(\omega l)^{-1}$. \footnote{ Detailed discussion of the high-frequency (shot wave length) approximation can be found in \cite{Misner:1974qy}.}

The amplitude $\ts{a}$ is a "slowly changing" complex vector. The following gauge transformation
\be\n{gaug}
\tilde{a}_{\mu}= a_{\mu}+\Psi_{,\mu}\hh \Psi={1\over \omega}\psi \exp(iS)\, ,
\ee
preserves the form of  (\ref{anz}) and one has
\be\n{apsi}
\tilde{a}_{\mu}=a_{\mu}+ip_{\mu}\psi+{1\over \omega}\psi_{,\mu}\, .
\ee
Lorentz gauge condition ${\cal A}_{\mu}^{\ ;\mu}=0$  implies
\be \n{lor}
p^{\mu} a_{\mu}-{i\over \omega}a_{\mu}^{\ ;\mu}=0\, .
\ee
The gauge transformation (\ref{gaug}) preserves this condition provided the following relation is valid
\be  \n{preser}
- \ts{p}^2\psi+{i\over \omega} [2p_{\mu}^{\ ;\mu}\psi +p^{\mu}\psi_{,\mu}] +{1\over \omega^2}\psi_{;\mu}^{\ ;\mu}=0
\ee
The field strength $\ts{\cal F}$ is
\ba
&&{\cal F}_{\mu\nu}=i\omega {\cal Z}_{\mu\nu} e^{iS}\hh {\cal Z}_{\mu\nu}={\cal B}_{\mu\nu}-{i\over\omega}{\cal C}_{\mu\nu} \, ,\n{fz}\\
&&{\cal B}_{\mu\nu}=p_{\mu} a_{\nu}-p_{\nu} a_{\mu}\hh {\cal C}_{\mu\nu}=a_{\nu ;\mu}-a_{\mu ;\nu}\, .\n{f3}
\ea
One can  show that
\ba \n{FEQ}
&&{\cal F}_{\mu\nu}^{\ \ ;\nu}=-\omega^2 j_{\mu} e^{iS}\, ,\\
&&j_{\mu} = {\cal B}_{\mu\nu}p^{\nu}-{i\over \omega}[ {\cal B}_{\mu\nu}^{\ \ ;\nu}+
{\cal C}_{\mu\nu}p^{\nu}] -{1\over \omega^2}{\cal C}_{\mu\nu}^{\ \ ;\nu}\, .
\ea
The potential (\ref{anz}) satisfies Maxwell equations if $j_{\mu}=0$.

Finally, let us discuss conditions imposed on the field  by the requirement that it is self or anti-self dual. For a self-dual field these conditions are
\ba
&& {\cal{Z}}_{\mu\nu} m^{\mu} n^{\nu}=0\, ,
\ \mbox{for\ } a=0
\, ,\n{ep0}\\
&&{\cal{Z}}_{\mu\nu}(-l^{\mu} n^{\nu}+\bar{m}^{\mu} m^{\nu})=0\, ,
\  \mbox{for\ } a=1
\, ,\n{ep1}\\
&&{\cal{Z}}_{\mu\nu} l^{\mu} \bar{m}^{\nu}=0\, ,
\  \mbox{for\ } a=2\, .\n{ep2}
\ea
For anti-self-dual field similar conditions can be obtained from these relations if one changes $\ts{m}\leftrightarrow \bar{\ts{m}}$ keeping $\ts{\cal Z}$ unchanged. In other words, if one found a self-dual solution of the form (\ref{fz}), then by taking a complex conjugation of $\ts{\cal Z}$ in this solution one gets an anti-self-dual solution. Relations (\ref{fz})--(\ref{f3}) imply that this operation is equivalent to change $\ts{a}\leftrightarrow \bar{\ts{a}}$ and $\omega \to -\omega$ in relations (\ref{fz})--(\ref{f3}). In particular, this means that when one uses the high-frequency expansion of the field equations,  only the terms of the odd power in $\omega$ are sensitive to the state of polarization of the field.

\subsection{Truncated equations}

In what follows we use expansions of different objects in powers of $1/\omega$. We use the following notation
\be
X\rav{n} \tilde{X}
\ee
to indicate that the quantities $X$ and $\tilde{X}$ differ only by terms of the order $O(\omega^{-(n+1)})$. Suppose that some relation $X=0$ depends on $\omega$ and  $X$ has a high-frequency expansion
\be
X=\sum_{k=0}^{\infty} {X_k \over \omega^k}\, .
\ee
If we keep the first $(n+1)$-terms in this expansion
\be
X^{(n)}=\sum_{k=0}^{n} {X_k \over \omega^k}\, ,
\ee
then $X^{(n)}\rav{n}0$. We call this relation a  $n$-th order truncated form of the equation $X=0$.

Using expressions  (\ref{f3}) for ${\cal B}_{\mu\nu}$ and ${\cal C}_{\mu\nu}$, the Lorentz condition (\ref{lor}) and keeping the terms up to the order $1/\omega$ one obtains
\be
j_{\mu} \rav{1} -\ts{p}^2 a_{\mu} +{i\over \omega} \left(
2  a_{\mu ;\nu}p^{\nu}+p^{\nu}_{\ ;\nu} a_{\mu}\right)\, .
\ee
Hence, the truncated field equations take the form
\be \n{ppaa}
\ts{p}^2 a_{\mu} -{2i\over \omega} \left(
 a_{\mu ;\nu}p^{\nu}+{1\over 2}p^{\nu}_{\ ;\nu} a_{\mu}\right) \rav{1} 0\, .
\ee

We denote $f^2= (\ts{a},\bar{\ts{a}})$ and write the complex amplitude $\ts{a}$ in the form
\be
a_{\mu}=f z_{\mu}\hh (\ts{z},\bar{\ts{z}})=1\, .
\ee
Then the equation (\ref{ppaa}) takes the form
\be \n{feq1}
\ts{p}^2 z_{\mu} -{2i\over \omega} [
z_{\mu ;\nu}p^{\nu}+( (\ts{q},\ts{p})+{1\over 2}p^{\nu}_{\ ;\nu}) z_{\mu} ] \rav{1} 0\, .
\ee
We call this relation a first order truncated field equation. It is sufficient for our purpose. However, it is easy to find extra terms of the higher in $1/\omega$ powers and to obtain a higher order truncated field equations.

The Lorentz gauge condition (\ref{lor}) written in $\{ f, \ts{z}\}$ variables is
\be \n{lorz}
(\ts{p},\ts{z})-{i\over \omega}\left[
(\ts{q},\ts{z})+z^{\mu}_{\ ;\mu}\right]=0\, .
\ee
Let us denote
\be
J= {1\over 2}( \bar{a}_{\mu} j^{\mu}+{a}_{\mu} \bar{j}^{\mu})\, .
\ee
Then one has
\ba\n{JRAV}
&&J\rav{1} -f^2[\ts{p}^2 -{2\over \omega} b_{\mu}p^{\mu}]\, ,\\
&&b_{\mu}={i \over 2}  \left( \bar{z}^{\nu} z_{\nu ;\mu}-z^{\nu}\bar{z}_{\nu ;\mu}\right)\, .
\ea

Let us remind that all the above results were obtained for right-handed circularly polarized high-frequency waves. One can easily repeat the calculations for the case of left-handed circularly polarized waves. However, this is not necessary. Instead of this one can use the prescription described at the end of subsection~\ref{hfe}. In particular, this means that  relation (\ref{JRAV}) can be used to get a similar relation for the left circular polarization. It is sufficient to take its complex conjugation and change $\omega\to -\omega$. Since $\bar{b}_{\mu}={b}_{\mu}$, the only change is the sign of the second term in the right-hand side of (\ref{JRAV}).
Hence the field equations $j_{\mu}=0$ in the both cases, $s=\pm 1$, imply
\be\n{disp}
\ts{p}^2 -{2s\over \omega} b_{\mu}p^{\mu}\rav{1} 0\, .
\ee
We call this  relation a dispersion equation.

In order to develop both geometric and spin optics approximations we use the first order truncated field and dispersion equations, (\ref{feq1}) and (\ref{disp}), and Lorentz condition (\ref{lorz}). We also add to them  first order truncated polarization equations (\ref{ep0})-(\ref{ep2}).  In this paper we restrict ourself by studying high-frequency solutions of the Maxwell equations in the first order approximation. However, in both geometric and spin optics approaches one can easily derive equations in the higher order approximation. Let us also remark that in order to obtain the first order truncated  polarization equations (\ref{ep0})-(\ref{ep2}) it is sufficient to substitute in them instead of the tensor $\ts{{\cal C}}$   its zero order approximation.

\section{Geometric optics}

A starting point of both, geometric optics and spin optics approximations is the same. Namely, one uses the first order truncated field equations (\ref{feq1}), (\ref{disp}) and the Lorentz gauge condition (\ref{lorz}). The difference between these approaches is in the procedure used for solving these equations. We describe the spin optics approach in the next section.
In this section we briefly remind the main steps of the standard geometric optics approximation\footnote{Additional details of the standard geometric optics approach to the Maxwell field propagating in a curved spacetime can be found \cite{Misner:1974qy}}.

\subsection{Effective Hamiltonian}

We start with the first order truncated equation (\ref{disp}).
Equating  to zero the lowest order in $\omega$ term in this equation  one gets
\be \n{pp0}
\ts{p}^2=0\, .
\ee
This equation shows that $\ts{p}$ is a null vector. It also implies that
\be \n{par}
0=(p_{\nu}p^{\nu})_{\mu}=2p^{\nu}p_{\nu ;\mu}=2p^{\nu} p_{\mu ;\nu}\, .
\ee
Here we used the property $p_{\mu ;\nu}=p_{\nu ;\mu}$.

Let $x^{\mu}(\lambda)$ be an integral line of $p^{\mu}$:
\be\n{xp}
\dot{x}^{\mu}\equiv {dx^{\mu}\over d\lambda}=p^{\mu}\, .
\ee
then
\be\n{DDD}
{D^2 x^{\mu}\over D\lambda^2}=0\, .
\ee
In other words,  $x^{\mu}$ is a null geodesic and $\lambda$ is an affine parameter.
We identify $\dot{\ts{x}}$ with a tangent vector $\ts{l}$ of the congruence of null rays.
 Since the acceleration parameter vanishes, $\kappa=0$, the null frame $(\ts{l},\ts{m},\bar{\ts{m}},\ts{n})$ associated with these rays is parallel transported along the rays. This frame is uniquely defined provided it is fixed at some initial moment of time.

The above results admit a slightly different but very useful interpretation. The relation $P_{\mu}=S_{,\mu}$ defines momenta  which are canonically conjugated to $x^{\mu}$. Let us consider an eight dimensional phase space with canonical coordinates $(x^{\mu},P_{\mu})$ and let
\be
\ts{\Omega}= dP_{\mu}\wedge dx^{\mu}\,
\ee
be a canonical symplectic  form in it. As usual, a summation over the repeated indices is assumed.
Let us write relation (\ref{xp}) in the form
\be \n{xP}
{dx^{\mu}\over d\lambda}={1\over \omega} g^{\mu\nu} P_{\nu}\, .
\ee
One can  introduce a Hamiltonian
\be
H={1\over 2\omega}g^{\mu\nu}P_{\mu}P_{\nu}\, ,
\ee
then (\ref{xP}) is identical with the first set of Hamiltonian equations
\be
{dx^{\mu}\over d\lambda}={\partial H\over \partial P_{\mu}}\, .
\ee
Using this equation together with the second set of Hamiltonian equations
\be
{dP_{\mu}\over d\lambda}=-{\partial H\over \partial x^{\mu}}
\ee
one obtains\footnote{For details see discussion in section~\ref{qqq}.}
\be\n{dxl}
{D^2 x^{\mu}\over D\lambda^2}=0\, .
\ee
As expected, this equation correctly reproduces (\ref{DDD}).

The Lagrangian $L$ of this system is
\be
L=P_{\mu}\dot{x}^{\mu}-H={\omega\over 2} \ts{\dot{x}}^2\, .
\ee
Both, the Hamiltonian and the Lagrangian do not depend on the polarization state  and trajectories of massless  particles with spin (photons) in the geometric optics approximation do not depend on their helicity.

\subsection{Polarization vector and amplitude}

Substituting equation (\ref{pp0}) into (\ref{feq1}) one gets
\be \n{zmnu}
z_{\mu ;\nu}p^{\nu}-[( \ts{q},\ts{p})+{1\over 2}p^{\nu}_{\ ;\nu}] z_{\mu}\rav{0}0 \, .
\ee
Multiplying this equation by $\bar{z}^{\mu}$ one gets
\be
( \ts{q},\ts{p})+{1\over 2}p^{\nu}_{\ ;\nu}\rav{0}\bar{z}^{\mu}z_{\mu ;\nu}p^{\nu}\, .
\ee
The quantity in the right hand side is purely imaginary. Really
\be\n{rez}
\Re (\bar{z}^{\mu}z_{\mu ;\nu})={1\over 2} (\bar{z}^{\mu}z_{\mu ;\nu}+{z}^{\mu}\bar{z}_{\mu ;\nu})=
{1\over 2}(\ts{z},\bar{\ts{z}})_{;\nu}=0\, .
\ee
Thus one has
\be \n{parq}
( \ts{q},\ts{p})+{1\over 2}p^{\nu}_{\ ;\nu}=0\, ,
\ee
while equation (\ref{zmnu}) gives
\be \n{parz}
z_{\mu ;\nu}p^{\nu}=0\, .
\ee
The Lorentz condition (\ref{lorz}) implies
\be \n{lor0}
(\ts{p},\ts{z})\rav{0}0\, .
\ee

We use the following expansions
\be
f\rav{1} f_0+{1\over \omega}f_1\hh
z^{\mu}\rav{1} z_0^{\mu}+{1\over \omega}  z_1^{\mu}\, .
\ee
Since $\ts{p}=\ts{l}$  relations (\ref{parq}) --(\ref{lor0})  imply
\ba
&&l^{\mu} z_{0 \mu}=0\, ,\n{llz}\\
&&l^{\nu}z^{\mu}_{0 \ ;\nu}=0\, ,\n{llzz}\\
&&l^{\mu} f_{0;\mu}=-{1\over 2} l^{\mu}_{\ ;\mu}f_0\, . \n{llf}
\ea
This means that the normalized amplitude vector $\ts{z}_0$ is parallel transported along the null rays and it is orthogonal to them. Equation (\ref{llf}) is a standard transport equation relating the change of the field amplitude with expansion of the null ray congruence. In what follows we shall use a slightly different form of this equation. We denote $q_{0 \mu}=\nabla_{\mu} \ln(f_0)$. Then (\ref{llf}) gives
\be\n{qqll}
(\ts{q}_0,\ts{l})+{1\over 2} l^{\mu}_{\ ;\mu}=0\, .
\ee

In the leading order polarization equation (\ref{ep2}) is identically satisfied, while the other two equations, (\ref{ep0}) and (\ref{ep1}),  give
\be
(\ts{z}_0,\ts{m})=(\ts{z}_0,\ts{l})=0\, .
\ee
Hence $\ts{z}_0=c_1 \ts{l}+c_2 \ts{m}$. Since $(\ts{z}_0,\bar{\ts{z}}_0)=1$ one has $c_2=e^{i\phi}$.

Under  gauge transformation (\ref{apsi}) with $\psi\rav{0}\psi_0 $
the zero-order term of the complex amplitude changes as follows
\be
\tilde{a}_{0 \mu}\rav{0}a_{0 \mu}+i\psi_0 l_{\mu}\, .
\ee
Since $\ts{p}^2=0$ the gauge transformation (\ref{gaug}) with arbitrary $\psi_0$ preserves the Lorentz condition. It can be used to put $c_1=0$. The parameter $\phi$ can be absorbed in a redefinition of the phase function.
Hence, one can put $\ts{z}_0=\ts{m}$.

Collecting all the above results one can write
\be\n{A0}
\ts{\cal A}^{+1}=f_0 \ts{m} e^{iS} \, .
\ee
Let us emphasize that both the equation for null rays, (\ref{dxl}) and the transport equations (\ref{llz})-(\ref{llf}) do not depend on the frequency $\omega$. This means that in order to obtain an anti-self-dual solution one can simply change $\ts{m}$ by $\bar{\ts{m}}$ in (\ref{A0})
\be
\ts{\cal A}^{-1}=f_0 \bar{\ts{m}} e^{iS} \,  .
\ee
Let us emphasize that  the  phase functions $S$ in both expressions for $\ts{\cal A}^{\pm1}$ are the same.
The fields with $s=+1$ and $s=-1$ describe right and left-handed circularly polarized waves, respectively \cite{Misner:1974qy}.

\section{Spin-optics}

\subsection{Effective Hamiltonian}

In the spin-optics approximation we use the same ansatz for the complex potential (\ref{anz}) as earlier.
However, do not require that the phase functions are the same for both polarizations. We only assume that their difference is small.  It is also convenient to present the scalar amplitude $f$ in the form $f=\exp(q)$.
Our starting point for construction of the spin optics approximation is again a truncated equation (\ref{disp}). But we proceed differently than in the  geometric optics case. First we add $\ts{b}^2/\omega^2$ term to its left-hand side. It is clear that this operation does not affect the truncated first order equation. Next we define an effective Hamiltonian $H$ by the relation
\be
H={1\over 2\omega} (\ts{P}-s\ts{b})^2\, .
\ee
In section~\ref{qqq} we discuss the Hamiltonian equations for this Hamiltonian and show that they are equivalent to
the following equations
\ba
&& \dot{x}^{\mu}={1\over \omega}(P^{\mu}-sb^{\mu})\, ,\n{pleq}\\
&&{D^2 x^{\mu}\over D\lambda^2}={s\over \omega} k^{\mu}_{\ \nu} \dot{x}^{\nu}\, ,\n{eqm}\n{RAY}\\
&& k_{\mu\nu}=b_{\nu ;\mu}-b_{\mu ;\nu}\, .\n{DD1}\n{rhs}
\ea
This system of equations  is invariant under the transformation
\be
\ts{b}\to-\ts{b}\hhh s\to -s\hhh \omega\to -\omega\hhh \lambda\to -\lambda\, .
\ee

Since $\ts{k}$ is antisymmetric, one has $d(\dot{\ts{x}})^2/d\lambda=0$. This means that if $\dot{\ts{x}}$ is null at some moment of time it remains null along the whole ray. The value of $H$ restricted to such rays is zero. Equation (\ref{RAY}) shows that if $\ts{k}\ne 0$ these null rays are not geodesics. For the congruence of these null rays we write  $l^{\mu}= \dot{x}^{\mu}$ and  introduce the associated null frame as it was described in section~\ref{NFR}.
The vectors of this frame are $F$-transported along the null rays and obey equations (\ref{NNN}).

\subsection{Polarization vector and amplitude}

We consider first the case of a right-handed circular polarization waves.
We use  relation $\ts{p}=\ts{l}+{1\over \omega}\ts{b}$ and write $q$ and $\ts{z}$ in the form
\be\n{exp}
q\rav{1}q_0 +{1\over \omega}q_1\hhh
\ts{z}\rav{1}\ts{m}+{1\over \omega}\ts{z}_1\, .
\ee
It is easy to see that for this choice all zero order truncated equations (\ref{disp}),  (\ref{feq1}), (\ref{lorz}) and polarization equations (\ref{ep0})-(\ref{ep2}) are satisfied. Really, the dispersion relation implies that $\ts{p}^2\rav{0} 0$, so that the leading zero order term in (\ref{feq1}) identically vanishes. The same is true for the Lorentz condition (\ref{lorz}) since $(\ts{p},\ts{z})\rav{0} 0$. As for zero order truncated polarization equations (\ref{ep0})-(\ref{ep2})
it is sufficient to omit the terms containing $\ts{\cal C}$ in them and to use for $\ts{\cal B}$ the following expression
\be
{\cal B}_{\mu\nu}\rav{0} l_{\mu} m_{\nu}-l_{\nu} m_{\mu}\, .
\ee
This means that in the zero order approximation $\ts{\cal B}$ coincides with the polarization tensor $\ts{\pi}^{(2)}$ determined by equation (\ref{polt}), and hence $\ts{\cal B}\circ \bar{\ts{\pi}}^{(a)}\rav{0}0$.

We consider now the first order truncated equations.
Let us substitute the truncated dispersion equation (\ref{disp}) into (\ref{feq1}). The leading term proportional to $\omega^0$ vanishes, while the first order term gives
\be \n{zpq}
m_{\mu ;\nu} l^{\nu}+[ (\ts{q}_0,\ts{l})+{1\over 2} l^{\mu}_{\ ;\mu}+ i(\ts{l},\ts{b}_0)] m_{\mu}\rav{0}0\, ,
\ee
where
\be
b_{0 \mu}={i\over 2}(\bar{m}^{\nu} m_{\nu ;\mu}-{m}^{\nu} \bar{m}_{\nu ;\mu})\, .
\ee

Multiplying this equation by $\bar{m}^{\mu}$ and using the property
$\Re(\bar{m}^{\mu} m_{\mu ;\nu})=0$
 one gets a relation
\be\n{qp0}
(\ts{q}_0,\ts{l})+{1\over 2} l^{\mu}_{\ ;\mu} =0\, ,
\ee
which is the same as (\ref{qqll}). This is a transport equation which determines the evolution of the scalar amplitude $f_0$ along the rays. After substituting (\ref{qp0}) into (\ref{zpq}) one obtains the following relation
\be
l^{\nu} m_{\mu ;\nu}+i(\ts{l},\ts{b}_0) m_{\mu}\rav{0}0\, .
\ee
This relation is valid because  $l^{\nu} m_{\mu ;\nu}\rav{0}0$. The last equation directly follows from (\ref{www}).
The above relations show that the first order truncated field equations are satisfied if the dispersion relation is valid, provided the scalar amplitude of the field obeys the transport equation (\ref{qp0}).

It is easy to check that in the first order the Lorentz condition (\ref{lorz}) is satisfied. Collecting the terms proportional to $\omega^{-1}$ in this relation one gets
\be \n{lorr}
(\ts{z}_1,\ts{l})+(\ts{b}_0,\ts{m}) -i [ (\ts{q}_0,\ts{m})+m^{\mu}_{\ ;\mu}]\rav{0}0\, .
\ee
One can simplify this relation using the property
\be
(\ts{b}_0,\ts{m})-im^{\mu}_{\ ;\mu}\rav{0} i l^{\mu}n^{\nu} m_{\mu ;\nu}\, .
\ee
Hence (\ref{lorr}) takes the form
\be\n{z1l}
(\ts{z}_1,\ts{l})=i\left[ (\ts{q}_0,\ts{m})-l^{\mu}n^{\nu} m_{\mu ;\nu} \right]\, .
\ee

Let us consider now the truncated polarization equations (\ref{ep0})--(\ref{ep2}). As we already mentioned, these equations are identically valid in the zero order approximation. Collecting the first order terms one obtains the following relations
\be \n{QQ}
\ts{\cal Q}\circ \bar{\ts{\pi}}^{(a)}=0\, ,
\ee
where
\ba
{\cal Q}_{\mu\nu}&=&(b_{\mu}m_{\nu}-b_{\nu}m_{\mu})+(l_{\mu}z_{1\nu}-l_{\nu}z_{1\mu})\nonumber\\
&-& i(m_{\nu ;\mu}-m_{\mu ;\nu})-i (q_{\mu}m_{\nu}-q_{\nu}m_{\mu})\, .
\ea
Rather long but straightforward calculations show that for $a=1$ and $a=2$ the relations (\ref{QQ}) are identically satisfied, while for $a=0$ one obtains
\be\n{z1m}
(\ts{z}_1,\ts{m})\rav{0}i m^{\mu}n^{\nu}m_{\nu ;\mu}\, .
\ee

Let us summarize. The first order truncated equations are satisfied if the scalar amplitude of the wave obeys the same transport equation (\ref{qp0}) as in the geometric optics case, while the normalized polarization vector $\ts{z}$ has the form (\ref{exp}) with a correction term $\ts{z}_1$ satisfying equations (\ref{z1l}) and (\ref{z1m}). These results can be adapted to the left-hand circular polarization case, $s=-1$. In this case one should put
\be
\ts{z}^{-1}=\bar{\ts{m}}-{1\over\omega} \bar{\ts{z}}_1\, .
\ee

\subsection{Equations of motion}

The equation of motion for circularly polarized rays (\ref{RAY}) can be further simplified. Let us notice that the right-hand side of this equations contains the factor $\omega^{-1}$. Hence keeping the same order one can put there $\ts{z}=\ts{m}$, so that
\be
b_{\mu}=i\bar{m}^{\alpha} m_{\alpha ;\mu}\, ,
\ee
and one gets
\ba
k_{\mu\nu}&=&b_{\nu ;\mu}-b_{\mu ;\nu}=i \bar{m}^{\alpha}(m_{\alpha;\nu\mu}-m_{\alpha;\mu\nu})\nonumber\\
&+&i(\bar{m}^{\alpha}_{\ ;\mu} m_{\alpha ;\nu}-\bar{m}^{\alpha}_{\ ;\nu} m_{\alpha ;\mu})\, .\n{KKK}
\ea
The term in the first brackets in the right hand side contains the commutator of the covariant derivatives and it is proportional to the curvature
\be
m_{\alpha;\nu\mu}-m_{\alpha;\mu\nu}=-R_{\mu\nu\beta \alpha}m^{\beta}\, .
\ee
Thus we have
\be\n{ekk}
k_{\mu\nu}=-iR_{\mu\nu \alpha\beta }m^{\alpha}\bar{m}^{\beta}
+i(\bar{m}^{\alpha}_{\ ;\mu} m_{\alpha ;\nu}-\bar{m}^{\alpha}_{\ ;\nu} m_{\alpha ;\mu})\, .
\ee
The right-hand side of (\ref{RAY}) contains the factor $k_{\mu\nu}l^{\nu}$. For the null frame $F$-transported along the null rays one has $l^{\nu}m_{\mu ;\nu}\rav{0}0$, so that the term in the  brackets in (\ref{ekk}) can be neglected.
Finally,  the polarized ray equation (\ref{RAY}) takes the following form
\be \n{NREQ}
{D^2 x^{\mu}\over D\lambda^2}=-{is\over \omega} {d{x}^{\nu}\over d\lambda} R^{\mu}_{\ \nu \alpha\beta }m^{\alpha}\bar{m}^{\beta} \, .
\ee
The left-handed side of this equation is nothing but the null ray acceleration $w^{\mu}$. Using relation (\ref{www}) one can find the acceleration parameter $\kappa$
\be \n{kap}
\kappa=-(\ts{w},\ts{m})=-{is\over \omega} R_{\mu\nu\alpha\beta} l^{\mu}m^{\nu} m^{\alpha} \bar{m}^{\beta}\, .
\ee

Let us remind that  in the derivation of the ray equation (\ref{NREQ}) we used the special ($F$-transported)  frame associated with the congruence of null rays, so that the following set of equations should also be  satisfied, (\ref{NNN}),
\be\n{NNN1}
  \dl \ts{n}=0\hhh \dl \ts{m}=-\kappa \ts{n}\hhh \dl \ts{\bar{m}}=-\bar{\kappa} \ts{n}\, ,
\ee
This set of equations guarantees that the proper normalization conditions for the null tetrad vectors are satisfied, provided they are valid at the initial moment. By solving the system of equations (\ref{NREQ})-(\ref{NNN1}) one obtains trajectories of the polarized  rays. Let us notice that since the rotation $\ts{m}\to \exp(i\varphi) \ts{m}$ preserves a two form $\ts{m}\wedge \bar{\ts{m}}$, this transformation also preserves the form of the equation (\ref{NREQ}). However, this equation is not invariant under the transformation $\ts{m}\to \ts{m}+a\ts{l}$ (see section~\ref{NF}).

In conclusion of this section, let us make a following remark.  Let us denote by $\ell$ the characteristic length of the curvature radius $\ts{R}\sim 1/\ell^2$. Then one can use this parameter $\ell$ to introduce dimensionless coordinates $\tilde{x}^{\mu}$,  affine parameter  $\tilde{\lambda}$ and the curvature $\tilde{\ts{R}}$ as follows
\be
\tilde{x}^{\mu}={x}^{\mu}/\ell\hhh\tilde{\lambda}={\lambda}/\ell \hhh\tilde{\ts{R}}=\ell^2\ts{R}\, .
\ee
Then the equation (\ref{NREQ}) written in these dimensionless variables takes the form
\be \n{NREQ1}
{D^2 \tilde{x}^{\mu}\over D\tilde{\lambda}^2}=-is\varepsilon {d\tilde{x}^{\nu}\over d\tilde{\lambda}}  \tilde{R}^{\mu}_{\ \nu \alpha\beta }m^{\alpha}\bar{m}^{\beta} \, .
\ee
Here $\varepsilon=(\omega \ell)^{-1}$. This is a dimensionless ratio of the wavelength and the characteristic scale of the problem $\ell$. Thus the deflection of the rays from null geodesics is small, as it is expected.

\section{Effective action}

\subsection{Action and Euler-Lagrange equations}

Let $\ts{l}$ be a null ray congruence and $\ts{m}$ and $\bar{\ts{m}}$ be two complex null vectors
which are properly normalized and $F$-transported along the rays.
Let us consider the following action
\be \n{act}
S={1\over 2}\omega\int \eta\ \dot{\ts{x}}^2 d\lambda +s \int (\ts{b},\dot{\ts{x}}) d\lambda\, .
\ee
This is a relativistic version of the action discussed in the paper \cite{Duval:2005ky}.
The 1-form $\ts{b}$ which enters the action (\ref{act}) is
\be
b_{\mu}={i\over 2}(\bar{m}^{\nu} m_{\nu ;\mu}-{m}^{\nu} \bar{m}_{\nu ;\mu})\, .
\ee
The normalization condition $(\ts{m},\bar{\ts{m}})=1$ allows one to write this quantity in the following equivalent form
\be
b_{\mu}=i\bar{m}^{\nu} m_{\nu ;\mu}\, .
\ee

This action is a functional of the world line $x^{\mu}(\lambda)$ and a Lagrange multiplier $\eta(\lambda)$.
As earlier we use the notations
\be
\dot{x}^{\alpha}={dx^{\alpha}\over d\lambda}\hhh \dot{\ts{x}}^2=g_{\mu\nu}\dot{x}^{\mu}\dot{x}^{\nu}\hhh
(\ts{b},\dot{\ts{x}})=b_{\mu} \dot{x}^{\mu}\, .
\ee
The action (\ref{act}) is invariant under reparameterization $\lambda\to \lambda'$ provided $\eta $ transforms as follows $\eta'={d\lambda'\over d\lambda}\eta$.

A variation of this action with respect to the Lagrange multiplier $\eta$ gives
\be\n{null}
\dot{\ts{x}}^2=0\, .
\ee
This condition guarantees that the world lines $x^{\mu}(\lambda)$ which enter as the argument of the action obey a restriction (\ref{null}) and hence are null curves.

To obtain equations which arise as a result of variation of the world line $x^{\mu}(\lambda)$ it is convenient to use the method of covariant variations which is described in the book \cite{DeWitt:2011nnj}. Let us perform a local variation of the worldline  $x^{\alpha}\to x^{\alpha}+{\delta}x^{\alpha}$.
If $\phi(x)$ is a scalar field then
\be
\delta \phi(x)=\phi_{,\alpha}\delta x^{\alpha}\, .
\ee
One also has $\dot{\phi}=\phi_{,\alpha} \dot{x}^{\alpha}$.

Let $\ts{v}$ be a tensor field and $\ts{v}(x(\lambda))$ is its restriction on the ray $x^{\alpha}(\lambda)$.
Following DeWitt \cite{DeWitt:2011nnj} and using his notations we define a covariant variation of tensor $\ts{v}$ as follows
\be
\bar{\delta}   \ts{v}=\ts{v}_{;\alpha}\delta x^{\alpha}\, .
\ee
In particular, the covariant variation of the metric tensor vanishes, $\bar{\delta}   \ts{g}=0$. Covariant variations obey the Leibniz rule when applied to factors in a product.
Using the relation
\be
\bar{\delta}  (\dot{\ts{x}}^2)=2\dot{x}_{\alpha}{D \delta x^{\alpha}\over D\lambda}\, ,
\ee
one gets
\be
\int   \eta \bar{\delta}(\dot{\ts{x}}^2) d\lambda =-2\int {D\over D\lambda}\left(\eta {D{x}_{\alpha} \over D\lambda}
\right)
\delta x^{\alpha} d\lambda\, .
\ee

Let us calculate  the covariant variation of the second term of the action (\ref{act}):
\ba
&& \bar{\delta} \int b_{\mu} \dot{x}^{\mu} d\lambda=
\int [(\bar{\delta}b_{\mu}) \dot{x}^{\mu}+b_{\alpha}\bar{\delta}(\dot{x}^{\alpha})]d\lambda\, ,\nonumber\\
&&=\int [b_{\mu;\alpha}- b_{\alpha;\mu}]\dot{x}^{\mu}\delta x^{\alpha}d\lambda
=\int k_{\alpha\mu}\dot{x}^{\mu}\delta x^{\alpha}d\lambda\, .
\ea
Here we used the properties
\be
\bar{\delta}(\dot{x}^{\alpha})={d\delta x^{\alpha}\over d\lambda}\hh
{d\over d\lambda}b_{\alpha}=b_{\alpha,\mu}\dot{x}^{\mu}\, .
\ee
and performed an integration by parts.
Combining these results one obtains the equation
\be\n{DDD}
{D \over D\lambda}\left(\eta{D{x}^{\mu} \over D\lambda}\right)
={s\over \omega} k^{\mu}_{\ \nu} \dot{x}^{\nu}\, .
\ee
Using the freedom in the choice of the parameter $\lambda$, we can put $\eta=1$. Then $\lambda$ becomes the canonical parameter and equation (\ref{DDD})   coincides with (\ref{RAY}).

\subsection{Hamiltonian equations}

\n{qqq}

The Hamiltonian is defined as follows
\be \n{HP}
H=\dot{x}^{\mu} P_{\mu}-L\, ,
\ee
where
\be\n{PXD}
P_{\mu}={\partial L\over \partial \dot{x}^{\mu}}=\omega \dot{x}_{\mu}+sb_{\mu}\, .
\ee
Thus one has
\be
H={1\over 2\omega} (\ts{P}-s\ts{b})^2
\equiv {1\over 2\omega} g^{\alpha\beta} (P_{\alpha}-s b_{\alpha})(P_{\beta}-s b_{\beta})\, .
\ee
The Hamiltonian equations are of the form
\ba
\dot{x}^{\mu}&=&{\partial H\over \partial P_{\mu}}={1\over \omega}g^{\mu\nu} (P_{\nu}-sb_{\nu})\, ,\n{XH}\\
\dot{P}_{\mu}&=&-{\partial H\over \partial x^{\mu}}={s\over \omega} (P^{\nu}-sb^{\nu}){\partial b_{\nu}\over \partial x^{\mu}}\nonumber\\
&-&{1\over 2\omega}{\partial g^{\alpha\beta} \over \partial x^{\mu}}  (P_{\alpha}-s b_{\alpha})(P_{\beta}-s b_{\beta})\, \n{PX} .
\ea
Here $\dot{a}$ means a derivative of $a$ with respect to the parameter $\lambda$,  $\dot{a}=da/d\lambda$.
Let us demonstrate that these equations reproduce the Euler-Lagrange equations (\ref{RAY}).
Substituting expression (\ref{PXD}) for $\ts{P}$, into (\ref{PX})  one gets the following relation
\be\n{mid}
 {d\over d\lambda}(\omega g_{\mu\nu}\dot{x}^{\nu}+s b_{\mu})= {\omega\over 2} {\partial g_{\alpha\beta}\over \partial x^{\mu}} \dot{x}^{\alpha}\dot{x}^{\beta}+s{\partial b_{\nu}\over \partial x^{\mu}} \dot{x}^{\nu}\, .
\ee
To obtain this relation we used the equality
\be
{\partial g^{\alpha\beta}\over \partial x^{\mu}}=-g^{\alpha\lambda} g^{\beta\rho}{\partial g_{\alpha\rho}\over \partial x^{\mu}}\, .
\ee
Let us note that
\be
{d g_{\mu\nu}\over d\lambda} ={\partial  g_{\mu\nu}\over \partial x^{\alpha}}\dot{x}^{\alpha} \hhh
{d b_{\mu}\over d\lambda} ={\partial  b_{\mu}\over \partial x^{\alpha}}\dot{x}^{\alpha}\, .
\ee
Using these relations and collecting terms with factors $\omega$ and $s$ in (\ref{mid}) one gets
\be \n{GG}
\omega g_{\mu\nu} \left( {d\dot{x}^{\nu}\over d\lambda}+\Gamma^{\nu}_{\alpha\beta} \dot{x}^{\alpha}\dot{x}^{\beta}\right)=s k_{\mu\nu}\dot{x}^{\nu}\,  ,
\ee
where
\be
\Gamma^{\nu}_{\alpha\beta}={1\over 2}g^{\nu\gamma}\left(g_{\alpha\gamma ,\beta}+g_{\beta\gamma ,\alpha}-g_{\alpha \beta ,\gamma}\right)\, .
\ee
The equation (\ref{GG}) can be written in the form
\ba
&&{D^2 x^{\mu}\over D\lambda^2}={s\over \omega} k^{\mu}_{\ \nu} \dot{x}^{\nu}\, ,\n{DD}\\
&& k_{\mu\nu}=b_{\nu ;\mu}-b_{\mu ;\nu}\, .
\ea
It is easy to see that this equation correctly reproduces the polarized rays equation (\ref{RAY}) with $\ts{k}$ defined in (\ref{ekk}).

The phase function $S(x)$ which enters the field ansatz (\ref{anz}) can be found as a solution of the Hamilton-Jacobi equation
\be
H(\nabla S,x)={1\over 2\omega}g^{\mu\nu}(S_{,\mu}-s b_{\mu})(S_{,\nu}-s b_{\nu})=0\, .
\ee

The above described results can be presented in a slightly different form. Let us instead of the canonical momenta $P_{\mu}$ introduce a generalized momenta
\be
\Pi_{\mu}=P_{\mu}-s b_{\mu}\, ,
\ee
and define new Poisson brackets as follows
\be \n{MPB}
\{x^{\mu},x^{\nu}\}=0\hhh \{ x^{\mu},P_{\nu}\}=\delta^{\mu}_{\nu}\hhh
\{\Pi_{\mu},\Pi_{\mu}\}=s k_{\mu\nu}\, .
\ee
Then it is possible to show that the Hamiltonian equations for
\be
H={1\over \omega}g^{\mu\nu}\Pi_{\mu}\Pi_{\nu}
\ee
with a modified symplectic form defined by relations (\ref{MPB}) are equivalent to the original Hamiltonian equations
(\ref{XH})-(\ref{PX})\footnote{
For discussion of this subject and further references see e.g. \cite{PhysRevD.75.025027,2015PhLB..743..478D}}.

\subsection{Initial conditions}

In order to solve the system of equations (\ref{NREQ})--(\ref{NNN1}) for polarized  null rays one needs to complement this system with initial conditions, that is to make a choice of the null frame $(\ts{l},\ts{m},\bar{\ts{m}}, \ts{n})$ at some initial moment of time. Let us discuss this point.

We denote by $\Sigma$ a spacelike surface and let $\ts{u}$ be a timelike unit vector orthogonal to it. In the vicinity of $\Sigma$ we introduce synchronous (Gaussian normal) coordinates $(\tau,y^{i})$ in which
$\tau=0$ is an equation of $\Sigma$ and the metric is of the form
\be
ds^2=-d\tau^2 +g_{ij}(\tau,y^i) dy^{i} dy^j\hh
i,j=1,2,3\, .
\ee
We denote by  $h_{ij}(y^i)=g_{ij}(\tau=0,y^i)$  a 3-metric on surface $\Sigma$ induced by its embedding  into the four-dimensional spacetime.

Our ansatz for a high-frequency approximate solution for a polarized beam of light is
\be
\ts{\cal A}=f\ts{m} \exp(iS)\, .
\ee
Here $f$ and $S$ are functions of coordinates $x^{\mu}$ in the four-dimensional spacetime. Let us denote by
 $f^0$ and $S^0$ the value of these quantities on $\Sigma$
 \be
 f^{0}(y^i)=f(\tau=0,y^i)\hh  S^{0}(y^i)=S(\tau=0,y^i)\, .
 \ee
For a beam of light which has finite size and finite duration in time the function $f^0$ vanishes outside some finite domain  on $\Sigma$. We focus on the initial conditions which has this property.

Equation $S^0=C$ with some constant $C$ defines a two-dimensional surface on $\Sigma$ which can be identified with a wavefront for the light beam (see e.g. discussion in \cite{Misner:1974qy}). A set of these wavefronts for different values of $C$ foliates $\Sigma$. A three vector $\vec{P}$, $P_i=S^0_{,i}$,
is orthogonal to the wavefront. It coincides with the direction of the wave front propagation.

For  a local observer $\ts{u}$ we define a two-dimensional plane  which is orthogonal to both $\ts{u}$ and $\vec{P}$. Denote by $\ts{e}_1$ and $\ts{e}_2$ two unit mutually orthogonal vectors tangent to this plane such that the set of vectors $(\ts{e}_1,\ts{e}_2,\vec{P})$ is right-handed. Denote $\ts{m}^0=(\ts{e}_1+i\ts{e}_2)/\sqrt{2}$.
We assume that $\ts{m}^0(y^i)$ coincides with the initial value of the normalized polarization vector $\ts{m}$ on $\Sigma$, that is
\be
m_{\mu}|_{\Sigma}=(0,{m}^{0}_{ i})\, .
\ee
The vector $\ts{b}$ is defined as follows
\be
b_{\mu}=i \bar{m}^{\nu} m_{\mu ;\mu}=i\bar{m}^{\nu} (m_{\nu ,\mu}-\Gamma_{\nu\mu \rho} m^{\rho})\, ,
\ee
where $\Gamma_{\nu\mu \rho}$ are four-dimensional Christoffel symbols for the metric $g_{\mu\nu}$. Since the time component of $\ts{m}$ vanishes at $\Sigma$, it is easy to check that the spatial components of the vector $\ts{b}$ on $\Sigma$ can be written in the form
\be
b_i= i\bar{m}^{j} (m_{j ,i}-\gamma_{k i j} m^{k})\, ,
\ee
where
\be
\gamma_{k i j}={1\over 2}\left(
{\partial h_{ik}\over \partial y^j}+{\partial h_{kj}\over \partial y^i}-{\partial h_{ij}\over \partial y^k}
\right)\,
\ee
are  three-dimensional Christoffel symbols calculated for the metric $\ts{h}$.

Let us denote
\ba
&&\nu_{\mu}=(0,\nu_i) \hh \nu_i= S^0_{,i}+s b_i\, ,\\
&& l_{\mu}=u_{\mu}+{1\over (\ts{\nu}^2)^{1/2}} \nu_{\mu}\hh  n_{\mu}=u_{\mu}-{1\over 2} l_{\mu}\, .
\ea
The four-dimensional vectors $\ts{l}$ and $\ts{n}$ are null and obey the condition $(\ts{l},\ts{n})=-1$. They, as well as the complex null vectors $\ts{m}$ and $\bar{\ts{m}}$, are defined on $\Sigma$. Thus for a given value of the phase function $S^0$ on $\Sigma$ we constructed a null frame on this initial surface. We use this choice as the initial conditions for the set of equations (\ref{NREQ}), (\ref{NNN1}). We also choose the canonical parameter $\lambda$ for the null ray to vanish at the initial time (on $\Sigma$). Since $S_{,\mu}=\omega l_{\mu}+s b_{\mu}$ one has
\be \n{omeg}
\omega=-u^{\mu} S_{,\mu}\, .
\ee
 That is,  the parameter  $\omega$ is nothing but the frequency of the wave as measured by the observer $\ts{u}$ at the initial moment of time. Let us remind that the canonical parameter $\lambda$ is defined up to its rescaling $\lambda\to C\lambda$ where $C=$const along the ray.
The condition (\ref{omeg}) fixes this ambiguity.

There is still freedom connected with an ambiguity of choice of the vector $\ts{m}$ at the initial moment of time,
$m_{\mu}\to e^{i\varphi(y^i)} m_{\mu}$.  This transformation generates the following change of three-vector $b_i\to b_{i}-\varphi_{,i}$, and it can be absorbed in the redefinition of the initial phase of the beam $S^0\to S^0+s\varphi$. As we already mentioned, the equation (\ref{NREQ}) is invariant under this transformation.

Let us finally find the phase function $S(x^{\mu})$ which enters the high-frequency field ansatz (\ref{anz}).
One has
\be
{dS\over d\lambda}\equiv {dS(x^{\mu}(\lambda))\over d\lambda}=S_{,\mu}\dot{x}^{\mu}\, .
\ee
Since $S_{,\mu}=P_{\mu}$ one can use  (\ref{PXD}) and write
\be \n{ddSS}
{dS\over d\lambda}=\omega \dot{\ts{x}}^2+ s(\ts{b},\dot{\ts{x}})\, .
\ee
Along the null ray $\dot{\ts{x}}^2=0$. Using the equations of $F$-transport (\ref{NNN}) one can also conclude that
$l^{\mu} \bar{m}^{\nu}m_{\nu ;\mu}=0$, so that the second term in the right-hand side of (\ref{ddSS}) vanishes as well.
Thus $dS/d\lambda=0$. This means that the phase function $S(x^{\mu})$ is constant along null ray trajectories.
By solving the ray equations one can find coordinates $x^{\mu}$ of a point on the trajectory which starts at a point $y^i$ on $\Sigma$ and reaches $x^{\mu}$ at the value of the canonical parameter equal to $\lambda$, $x^{\mu}=x^{\mu}(\lambda,y^i)$.  Taking the inverse of these relations one gets
\be
\lambda=\lambda(x^{\mu})\hh y^i=y^i(x^{\mu})\, .
\ee
Since $S$ is constant along the rays one obtains
\be
S(x^{\mu})=S^0(y^i(x^{\mu}))\, .
\ee
This means that after the integration of the polarized ray equations one can restore the phase function in the field ansatz (\ref{anz}) by using its initial value $S^0$, and hence to obtain a required high-frequency approximate solution of the Maxwell equations in a curved spacetime.

\section{Discussion}

Let us summarize the obtained results. In order to describe propagation of high-frequency monochromatic beam of circularly polarized light in a curved spacetime one needs first to find a solution of the set of ordinary differential equations (\ref{NREQ})--(\ref{NNN1}). A choice of a beam is specified by imposing initial conditions on the null rays at some moment of time. After fixing a null frame associated with this beam at the initial time its propagation along the null rays is determined by equations (\ref{NNN1}).
Equation (\ref{NREQ}) shows that in the presence of curvature the motion of a circularly polarized photons is non-geodesic. A trajectory of such a photon depends both on its helicity and frequency.  A linearly polarized light can be presented as a superposition of the right and left-handed circularly polarized states with equal amplitudes.  This means that a  beam of light which is initially linearly polarized during its propagation in the gravitational field can split into two spatially separated circularly polarized beams with the opposite states of the helicity.

Another consequence of the equation (\ref{NREQ}) is the following. Suppose a distant observer registers time when burst of a circularly polarized light  emitted at some point reaches the point of the observation. In the limit
$\omega \to \infty$ the motion  of such photons is geodesic. However, if the frequency $\omega$ is finite this is not  anymore true. At the same time polarized photons still propagate with the speed of light. It is well known that it takes longer time for such photons to reach the point of observation. This statement is known as a generalized Fermat principle \cite{Perlick:1990,2000rofp.book.....P,Perlick:2006,Frolov:2013sxa}.
This effect of a time delay for circularly  polarized photons in a gravitational field is another important consequence of the spin optics equation (\ref{NREQ}). Let us emphasize that  this time delay depends both on their frequency and helicity. This opens an interesting principal opportunity for the observation of this effect.

The equation (\ref{NREQ}) is obtained in the spin optics approximation for  the propagation of the high-frequency electromagnetic waves in the curved spacetime so that the helicity parameter $s$ which enters it has values $\pm 1$. However, the form of this equation suggests that it should be valid for other massless fields with spin, in particular, for propagation of the gravitational waves. It is interesting to develop the spin optics approach to this case.

It should be emphasized that we often refer to spin optics in application to the polarized {\em light} propagation. But certainly this approach is applicable to all kinds of the electromagnetic waves including radio waves. The only limitation is that the corresponding wavelength is much smaller that characteristic length scale of the problem. It is interesting to search for possible observable polarization depended effects for electromagnetic and gravitational waves propagation in the cosmological and black hole backgrounds. In particular, it is well known that geodesic equations in the Kerr geometry are completely integrable. Is the same property valid for the polarized light equation (\ref{NREQ})? If  the presence of the curvature violates the complete integrability of this equation then the motion of polarized photons in the Kerr geometry may become  chaotic. In particular, this may affect the properties of a shadow of black holes.

\section*{Acknowledgments}
The author thanks the Natural Sciences and Engineering Research Council of Canada and the Killam Trust for their financial support. He  is also grateful to Andrei Frolov for many stimulating discussions.

\newpage

%

\end{document}